\documentclass[10pt,letterpaper]{article}

\usepackage{cogsci}

\cogscifinalcopy 

\usepackage{csquotes}
\usepackage{url}
\usepackage{graphicx}
\usepackage{booktabs}
\usepackage{multirow}
\usepackage{enumitem}
\usepackage{float} 
\usepackage{subcaption}
\usepackage{titlesec}
\usepackage{makecell}

\usepackage[
  style=apa,
  natbib=true,
  isbn=false,
  url=false,
  doi=false
]{biblatex}
\usepackage{hyperref}
\hypersetup{
    colorlinks,
    citecolor=black,
    filecolor=black,
    linkcolor=black,
    urlcolor=black
}

\titleformat{\paragraph}[runin]
  {\normalfont\normalsize\bfseries\itshape}  
  {\theparagraph}                             
  {0em}                                       
  {} 
\titlespacing{\paragraph}{0pt}{0.25ex plus 1ex minus .2ex}{0.3em}

\title{Why Someone Asked `Why': \\Foil Inference in Human and LLM Question Interpretation}

\author[1*]{\mbox{Britt Besch}}
\author[1]{\mbox{Tobias Gerstenberg}}
\affil[*]{\texttt{bb720@cam.ac.uk}}
\affil[1]{Department of Psychology, Stanford University}

\begin{document}

\maketitle

\begin{abstract}
Explanations are inherently contrastive: E happened rather than E' because of C rather than C'. However, these contrasts, or ``foils'', are rarely mentioned explicitly but have to be inferred in context. Here, we investigate how people select the intended foil E' of a why-question. 
Participants read vignettes and judged, for each foil, their prior expectation (what will happen next), closeness (what is most similar to what happened), and hindsight expectation (what could have happened instead), as well as which foil they thought the question asker had in mind when they asked the why-question. We found that foil selections were best predicted by hindsight expectation judgments. This suggests that people infer the foil by considering what a question asker finds surprising \textit{after} the outcome occurred. Since correct foil selection is relevant not only in human-human interaction but also increasingly in dialogues with large language models, we investigated their performance on the same task. The coupling between LLMs' explicit expectation judgments and their foil selections is inconsistent.


\end{abstract}

\section{Introduction}

When the bank robber Willie Sutton was asked ``Why did you rob banks?'', he famously answered ``because that's where the money is'' \citep{CongressionalRecord1995}. This response was unexpected.
His explanation answers a different why-question than the intended one. He explains why he robs banks rather than other places, whereas the question asker had wondered why he robs banks rather than not doing so. 

This example highlights the importance of inferring the correct, implicit contrast of a why-question in order to provide a helpful answer. Depending on which contrast the listener infers (``not rob the bank'' vs. ``rob other places''), their answer to the same question will be very different.


\subsubsection{The contrastive nature of explanations}
A large body of literature investigating the nature of explanation has explored what makes for a good explanation. Accordingly, good explanations are selective \citep{Lombrozo2007Simplicity, GerstenbergIcard2019Expectations, PoesiaReisSilvaGoodman2022}, causal \citep{HiltonErb1996MentalModels, Hilton1990Conversational} and contrastive \citep{lipton1990contrastive,Miller2021ContrastiveStructural}.

Experimental evidence confirms that humans naturally incorporate contrasts -- so-called ``foils'' -- into their explanations. For example, in a category-learning study, \textcite{ChinParkerCantelon2017} showed that when people were asked to explain why an item belonged to a certain category, their explanations implicitly referenced an alternative category as a contrast. This influence of an implicit foil also affected how well participants learned and transferred concepts \citep{ChinParkerCantelon2017,BERGEY2023105597}.

The contrastive nature of explanations is further linked to the contrastive structure of why-questions that take the form ``Why did $E$ happen instead of $E'$?''. A good answer to the question then explains the reason for why $E$ happened instead of $E'$ in the format of ``because $C$ instead of $C'$'' \citep{lipton1990contrastive,Schaffer2005ContrastiveCausation,Schaffer2010ContrastiveLaw}. 

\subsubsection{What is the foil of a why-question?}
In the opening example, the event $E$ is ``rob the bank''. From Willie Sutton's explanation we can infer that he implicitly contrasts $E$ with an $E'$ that can be summarized as something like ``other places''. His explanation then addresses this contrast: He robbed the bank ($E$) instead of other places ($E'$), because banks ($C$) instead of other places ($C'$) are where the money is. Note that if it is clear what the relevant foil of the question ($E'$) is, the contrast of the explanation ($C'$) often need not be spelled out explicitly to provide a good explanation. 
So, if we have $E$, $C$, and one of either $E'$ or $C'$ given, it is rather straightforward to infer the missing $E'$ or $C'$, respectively. In our example, however, only $E$ and $C$ were stated explicitly. Nonetheless, we were able to recognize that the $E'$ of the question asker and the answerer Willie Sutton differed. 
But how do we infer $E'$ in the first place? What made us think it was ``not robbing'' rather than ``other places''?

\subsubsection{Previous approaches on the nature of foils}
Previous literature on foils of why-questions is scarce. Work by \textcite{Hilton1990Conversational} and \textcite{van2002remote} distinguishes multiple contrast structures (e.g., property contrasts, object contrasts, temporal contrasts, norm contrasts). However, while these qualitative taxonomies theoretically characterize categories of possible contrasts, they leave open how an explainer infers which specific contrast is intended. 

\textcite{chin2017contrastive} argue that the relevant contrast class is inferred from context and show that changing background context alters which explanatory style is used \citep[see also][]{chin2010background}. However, which concrete contextual cues people use to identify that foil remains open. \textcite{debrigard2021perceived} show that people judge counterfactual worlds as more plausible when the imagined possible world in which they occur is perceived as more similar to the actual world. They conclude that perceived similarity is a key psychological driver of counterfactual plausibility. Even though this work does not directly address contrastive questions, its results indicate that the similarity between $E$ and $E'$ might influence the choice of $E'$.

Another promising approach stems from the fact that the reason a question is asked in the first place is that the questioner is surprised by event $E$ -- their prior expectation of what would happen has been violated \citep{chandra2024}. This suggests that the chosen $E'$ is related to the questioner's prior expectation \citep{Hilton1990Conversational}, or, more precisely, what the explainer believes the questioner's prior expectation was. 

Notice that for the questioner, their belief does not change before and after having asked the question (and before they have received an explanation). However, for the explainer the question carries further information that allows them to update their beliefs about a plausible $E'$ that the questioner has in mind. Namely, the explainer learns what event $E$ actually happened. Thus, it is plausible that an explainer infers $E'$ by considering the questioner's expectation in hindsight, after hearing the inquiry.

\subsubsection{Contrastive explanations and why-questions in LLM applications}

The growing interaction with AI systems, in particular LLMs, raises the question of how well these systems align with the kinds of inferences that people make from why-questions. \textcite{Miller2017ExplanationAI}, surveying the literature, noted that for developing explainable intelligent systems, perhaps the most important insight from the social sciences is that explanations are contrastive. \textcite{paranjape2021prompting} and \textcite{yao2024contrastive} demonstrate that prompting models to explicitly contrast correct and incorrect answers improves zero-shot accuracy on reasoning tasks. Contrastive prompting also helps LLMs generate explanations that are more aligned with human preferences. In retrieval-based settings, contrastive approaches like Contrastive-RAG \citep{gu2025core,ranaldi2025eliciting} train models to distinguish relevant from irrelevant evidence, improving performance on open-domain question answering. In human-computer interaction, contrastive explanations help users better understand model behavior and improve their own decision-making, especially when tailored to anticipated user misconceptions \citep{bucinca2025contrastiveexplanationsanticipatehuman}. However, this body of literature so far has only tried to elicit contrastive reasoning to generate better explanations. To our knowledge, there is no work on how well the contrastive interpretation of a \emph{question} aligns with how humans infer the foil when it is not explicitly stated.

\subsection{Hypotheses}

We investigate how people infer the foil in a why-question and compare this to the behavior of LLMs. We have the following hypotheses:

\begin{itemize}[leftmargin=0.75cm, itemsep=0cm]
\item[\textbf{H1}] ``prior expectation'': The more an event is expected to happen, the more likely it is to be selected as the foil. 

\item[\textbf{H2}] ``closeness'': The closer an event is to the one that has actually happened, the more likely it is to be selected as the foil. 

\item[\textbf{H3}] ``hindsight expectation'': The more an event was expected to have happened instead of the one that actually happened, the more likely it is to be selected as the foil.

\item[\textbf{H4}] Hindsight expectation is a better predictor of foil selection than prior expectation or closeness.
\end{itemize}


\section{Experiment 1: Humans}

\subsection{Methods}

\begin{figure*}[t]
  \centering
  \begin{subfigure}[t]{0.50\textwidth}
    \centering
    \includegraphics[width=\linewidth]{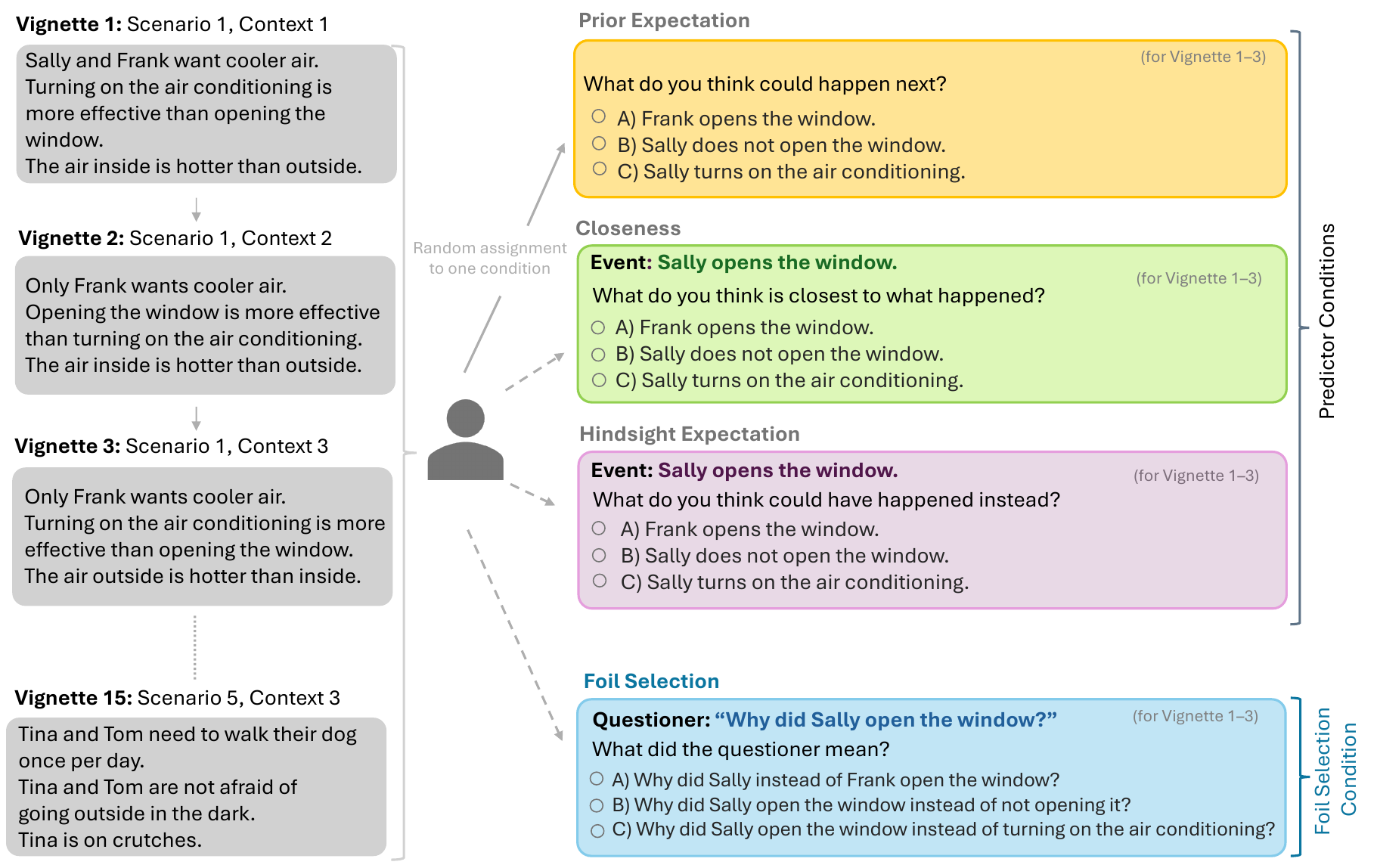}
    \caption{}
    \label{fig:scenario1_vignettes}
  \end{subfigure}
  \hfill
  \begin{subfigure}[t]{0.47\textwidth}
    \centering
    \includegraphics[width=\linewidth]{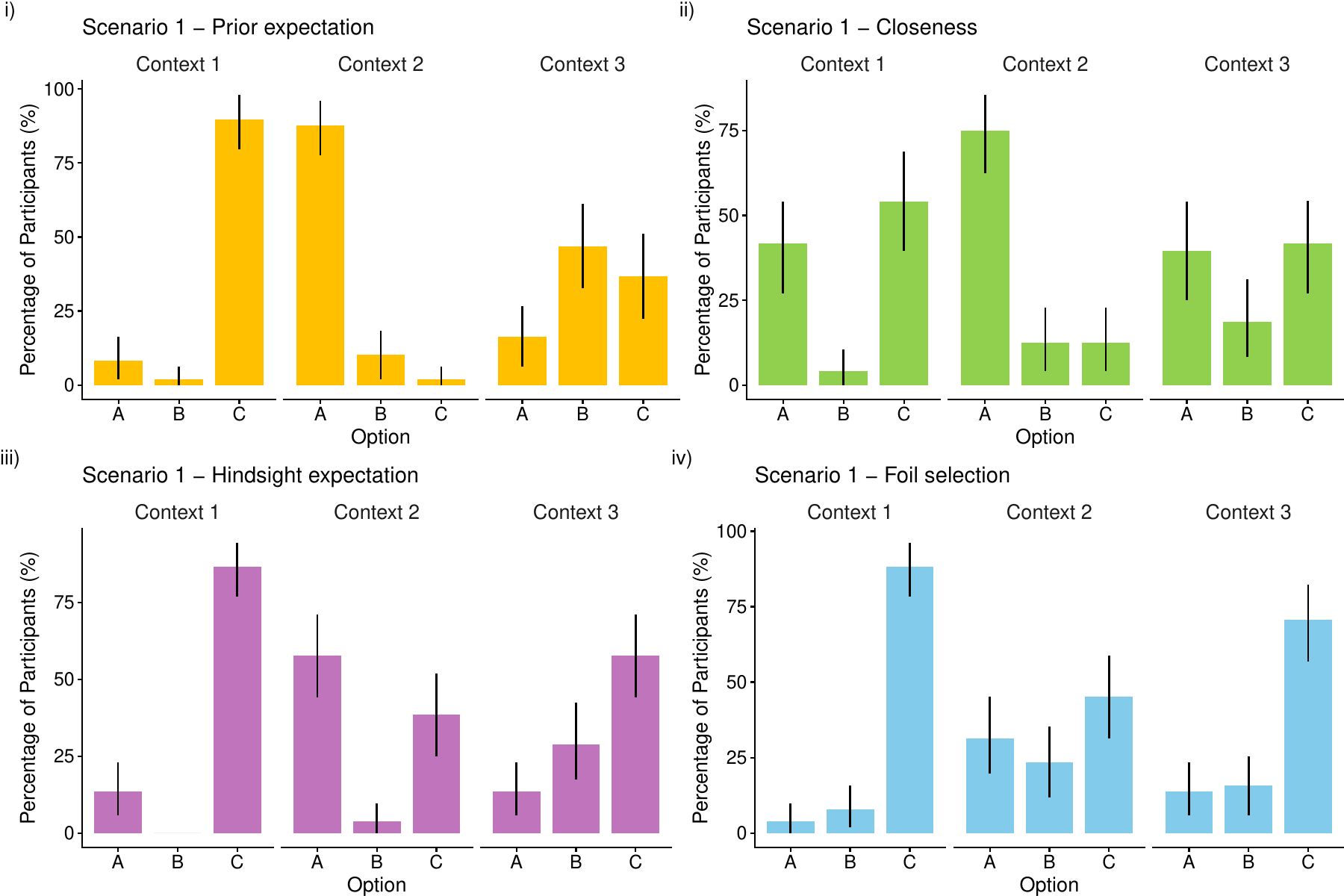}
    \caption{}
    \label{fig:scenario1_descriptive}
  \end{subfigure}
  \caption{\textbf{Experiment overview and Scenario~1 data}. (a) Experiment procedure. All participants read the same vignettes and were assigned to either one of the three predictor conditions (`prior expectation', `closeness', or `hindsight expectation') or the `foil selection' condition. 
  (b) Experiment 1 (Humans) results for Scenario~1 with three context variations. Bars show percentage of participants selecting each option. Error bars are 95\% bootstrapped confidence intervals (CIs).}
  \label{fig:overview}
  \end{figure*}

The study was pre-registered on the Open Science Framework and all materials, data and code are provided at: \url{https://github.com/cicl-stanford/why_someone_asked_why_CogSci26}.

\subsubsection{Participants}
Participants were recruited through Prolific and the experiment was run using \texttt{jspsych} \citep{deleuw2023jspsych}. Participants were paid \$12 an hour. Participants were required to participate using a desktop or laptop (no tablet or phone). Further inclusion criteria were that they were based in the US, fluent in English, have an approval rating of 97\% (or more) and at least 100 previous submissions.
We aimed at recruiting 50 participants for each condition so $N = 200$ in total. 10 participants dropped out during the experiment. We recruited 10 more participants. We ended up with $N = 51$, $N = 49$, $N = 48$, $N = 52$, for the conditions ``foil selection'', ``prior expectation'', ``closeness'', and ``hindsight expectation'', respectively (\emph{age}: median $= 43$, range = [19--78]; \emph{gender}: 104 female, 91 male, 2 non-binary, 1 queer/trans, 2 no response; \emph{race}:  168 White,  14 Black, 9 Asian, 9 Multiracial/Other).

\subsubsection{Vignettes}
We designed five different scenario-vignettes. We altered each scenario slightly, creating three variations per scenario. Each participant read and judged these 15 different vignettes.

Each vignette consisted of three sentences describing: the agent's action and goal, situational context such as environmental conditions or available resources, and agent preferences, capabilities, or properties of other agents. Response options included alternatives differing in: the way the action was performed, the action itself, the performing agent, the object of the action, or the action not occurring at all.
Example context variations of Scenario~1 are shown in \autoref{fig:scenario1_vignettes}, and the full set of vignettes is available in the online materials.

\subsubsection{Procedure}

\autoref{fig:overview} shows an overview of the experiment design. Participants were assigned to either one of the three predictor conditions or to the foil selection condition. The same 15 scenarios were presented in all conditions. However, a different question and three different answer options were presented to the participants in different conditions. We shuffled the order of the response options. Participants could choose one answer per vignette. We randomized the order in which vignettes were presented.

In the foil selection condition, participants were presented with a question about an event that occurred in the scenario and were asked to identify what the questioner meant (``What did the questioner mean?'') by selecting among different contrastive interpretations of the question. 
For example, for Scenario~1, participants were presented with the event \textit{``Sally opens the window.''} and the question \textit{``Why did Sally open the window?''}, and were asked to select among the options displayed in \autoref{fig:scenario1_vignettes}. They were instructed to assume the asker has the same knowledge of the scenario as they do.

In the prior expectation condition, participants were asked to indicate what they thought could happen next in the scenario (``What do you think could happen next?''), selecting from a set of possible events.
For Scenario~1, participants were presented with the same scenario contexts and asked to select from the options shown in \autoref{fig:scenario1_vignettes}. 
    
In the closeness condition, participants were presented with an event that occurred in the scenario and were asked to identify which alternative event was closest to what actually happened (``What do you think is closest to what happened?'').
For Scenario~1, participants were presented with the event \textit{``Sally opens the window.''} and asked to select from the same options as in the prior expectation condition.

In the hindsight expectation condition, participants were presented with an event that occurred in the scenario and were asked to indicate what they thought could have happened instead (``What do you think could have happened instead?'').
For Scenario~1, participants were presented with the event \textit{``Sally opens the window.''} and asked to select from the same options as in the prior expectation and closeness conditions. The event presented is always the same, but the question and answer options differ between conditions.

\subsection{Results}

\begin{figure}[b]
  \centering
  \includegraphics[width=\columnwidth]{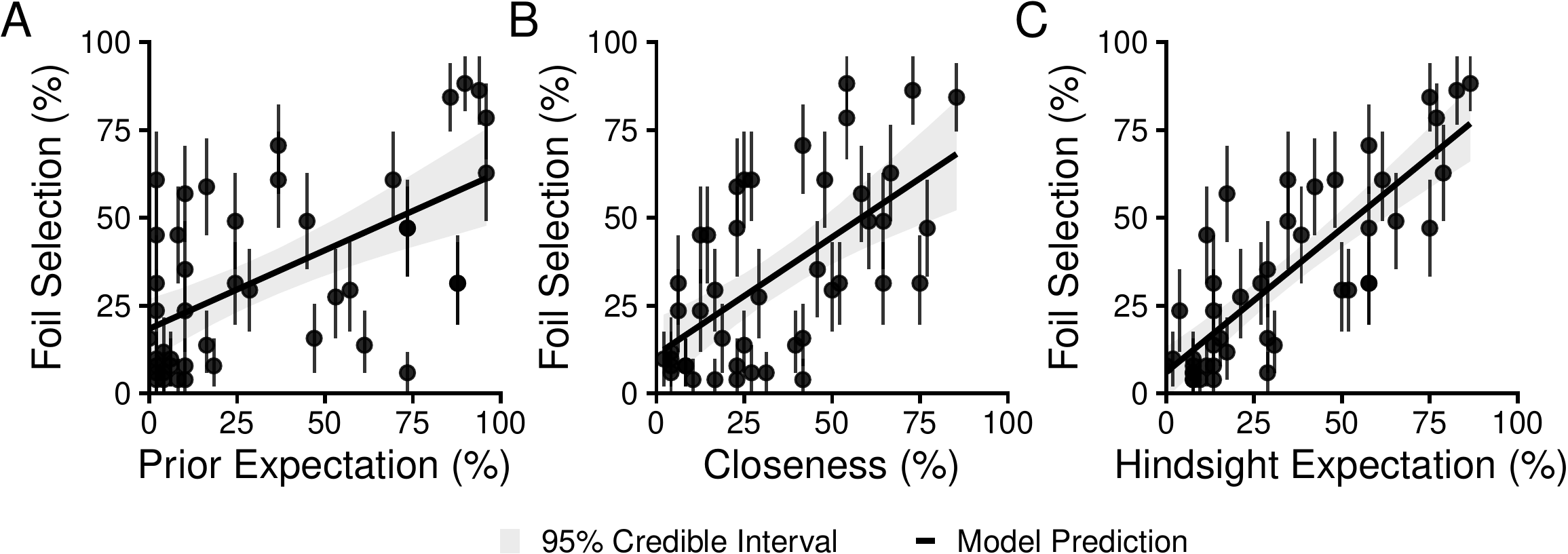}
  \caption{\textbf{Experiment~1 (Humans)}. Scatterplots of the percentage with which each response option of a vignette was selected in the predictor conditions, and the foil selection condition. Error bars are 95\% bootstrapped CIs.}
  \label{fig:scatterplots_human}
\end{figure}

The descriptive choice distribution of participants for scenario~1 in all four conditions is displayed as an example in \autoref{fig:scenario1_descriptive}. Plots for the rest of the scenarios can be found in the online materials. Across vignette $\times$ option percentages ($N = 45$), foil selection correlated with prior expectation ($r = .58$), closeness ($r = .62$), and most strongly with hindsight expectation ($r = .82$) (all $p < .001$). The predictors were also correlated with each other: $r(\text{prior}, \text{hindsight}) = .86$, $r(\text{prior},\text{closeness}) = .67$, $r(\text{hindsight}, \text{closeness}) = .73$, all $p < .001$.

We report two types of analyses: a preregistered Bayesian linear regression, and a Bayesian multinomial model.

\subsubsection{Linear regression}
To test our four hypotheses and investigate how well the three different predictors capture participants' foil selection, we fit four linear regression models. 
For each vignette we computed the percentage of participants choosing each option in each condition. We regressed foil-selection percentage (per option and vignette) on the corresponding predictor-condition percentages. Separate regressions tested H1 (foil $\sim$ prior expectation), H2 (foil $\sim$ closeness), H3 (foil $\sim$ hindsight expectation), and a combined model tested H4 (foil $\sim$ hindsight + prior + closeness). We fitted all models using \texttt{brms}. Hypotheses are accepted if the corresponding $\beta$ is positive and the credible interval (CrI) does not include zero. 

\begin{figure*}[t]
  \centering
  \includegraphics[width=\textwidth]{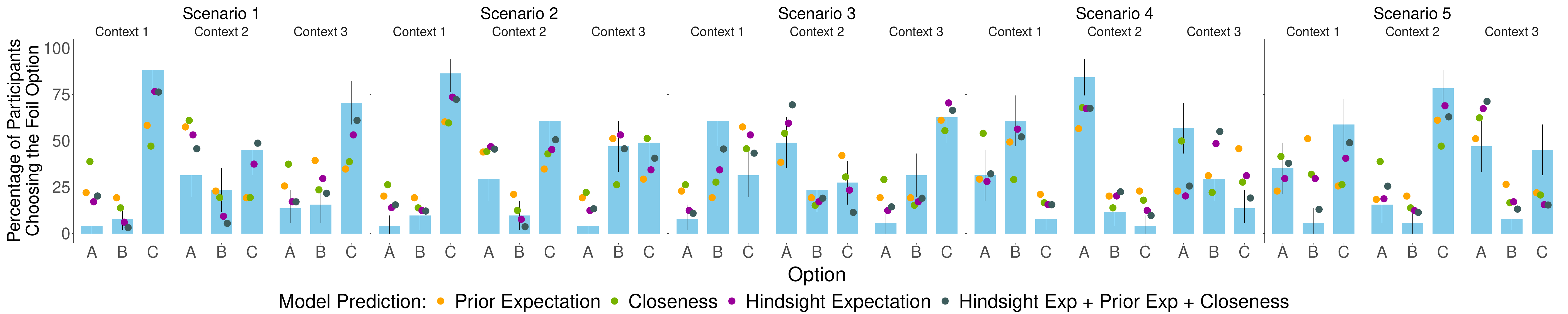}
  \caption{\textbf{Experiment 1 (Humans)}. Participant selections (bars) for all vignettes and model predictions of linear models with different predictors (points). Error bars on bars are 95\% bootstrapped confidence intervals.} 
  \label{fig:prediction_regression_percentages_human}
\end{figure*}

\autoref{fig:scatterplots_human} shows the scatterplots for foil selection percentages against predictor percentages for each vignette. \autoref{tab:exp1_primary} summarizes the regression coefficients, 95\% CrIs, and $R^2$ for all hypotheses. Thus, H1--H4 were supported. Each predictor was positively associated with foil selection (H1--H3), and hindsight was the best predictor (H4). \autoref{fig:prediction_regression_percentages_human} shows the prediction plots for all scenarios showing actual foil selection percentages (bars) and model predictions (points).

\begin{table}[t]
  \centering
  \caption{\textbf{Experiment 1 (humans)}. Linear regression of foil-selection predictor-conditions.}
  \label{tab:exp1_primary}
  \resizebox{\columnwidth}{!}{
  \begin{tabular}{@{}lccc@{}}
  \toprule
  Hypothesis & Predictor & $\beta$ [95\% CrI] & $R^2$ \\
  \midrule
  \multicolumn{1}{l}{H1} & Prior expectation & 0.45 [0.26, 0.63] & .33 \\
  \multicolumn{1}{l}{H2} & Closeness & 0.67 [0.40, 0.92] & .37 \\
  \multicolumn{1}{l}{H3} & Hindsight expectation & 0.82 [0.64, 0.99] & .67 \\
  \midrule
  \multicolumn{4}{l}{\textit{H4 (combined model)}} \\
  \quad Hindsight Expectation& & 1.16 [0.82, 1.52] & \\
  \quad Prior Expectation & & $-$0.37 [$-$0.62, $-$0.12] & \\
  \quad Closeness & & 0.10 [$-$0.16, 0.36] & \\
  \quad Model $R^2$ & & & .72 \\
  \midrule
  \multicolumn{4}{l}{\textit{Contrasts (H4):}} \\
  \quad $\Delta\beta$(hindsight $-$ prior) & & 1.53 [1.07, 1.99] & \\
  \quad $\Delta\beta$(hindsight $-$ closeness) & & 1.06 [0.64, 1.49] & \\
  \bottomrule
  \end{tabular}%
  }
  \end{table}

\subsubsection{Multinomial prediction}
Since percentages are bounded, heteroskedastic, and compositionally dependent within vignettes, we also ran a Bayesian multinomial model directly on participants' raw choices for robustness.
We fit three \texttt{brms} models (prior, closeness, hindsight) with $\text{choice} \sim 1 + (1|\text{vignette\_id})$ and evaluated expected log predictive density on the foil test set vs.\ a uniform baseline. Positive $\Delta\text{ELPD}$ values indicate that a predictor outperforms the baseline in predicting participants' foil choices, with a predictor considered reliable when $|\Delta\text{ELPD}|$ exceeds twice its SE.

Hindsight expectation showed the strongest predictive performance, substantially outperforming the uniform model ($\Delta\text{ELPD}=138.53$, SE = 19.91). Closeness also exceeded the uniform baseline ($\Delta\text{ELPD}=49.52$, SE = 19.43), whereas prior expectation did not ($\Delta\text{ELPD}=-104.37$, SE = 33.20).

\subsection{Discussion}

Hindsight expectation showed the strongest association with foil selection and dominated the combined model. Because the three predictors share substantial variance, coefficients in the combined model should be interpreted as conditional effects; under such multicollinearity, attenuation or sign reversals do not contradict the positive associations observed in single-predictor models. Importantly, hindsight expectation accounted for the largest share of variance in foil selection beyond what is shared with prior expectation and closeness. This is supported by the results of the multinomial analysis. Both analyses converge on the conclusion that people infer foils primarily from what they think \emph{could have happened instead}, rather than from what could happen next or what is most similar. Observing event $E$ seems to shift expectations. This is likely because the explainer's beliefs are updated through $E$, which allows for further understanding of the scene and additional inferences about it. Thus, some alternatives might become more or less salient than they were before having learned about $E$. For instance, in Vignette~3 (see \autoref{fig:overview}), before knowing that Sally opened the window, one might identify Frank as a potential actor (as he desires cooler air), and turning on the air conditioning as effective and opening the window as ineffective. This makes the scenario ambiguous, with both B and C, and to a lesser extent A, being salient. However, after learning that Sally opened the window, one could infer that she might also suffer from the hot air in the room or want to do Frank a favor. In combination with the knowledge about the scene, Sally as an actor and the effective action of turning on the air conditioning (option C) become the most likely $E'$.

\section{Experiment 2: Large Language Models}

\begin{table}[b]
  \centering
  \caption{\textbf{Experiment 2 (LLMs)}. Linear regression of foil-selection on predictor-condition percentages by LLM.}
  \label{tab:exp2_primary}
  \resizebox{\columnwidth}{!}{%
  \small
  \begin{tabular}{@{}llccc@{}}
  \toprule
  LLM & Hypothesis & Predictor & $\beta$ [95\% CrI] & $R^2$ \\
  \midrule
  \multirow{3}{*}{GPT-5.2} & H1 & Prior & 0.31 [0.06, 0.56] & .12 \\
   & H2 & Closeness & $-$0.23 [$-$0.52, 0.07] & .06 \\
   & H3 & Hindsight & 0.55 [0.34, 0.75] & .37 \\
  \midrule
  \multirow{3}{*}{Llama-3.1-70B} & H1 & Prior & 0.33 [0.03, 0.64] & .11 \\
   & H2 & Closeness & 0.17 [$-$0.16, 0.51] & .04 \\
   & H3 & Hindsight & 0.55 [0.34, 0.75] & .31 \\
  \midrule
  \multirow{3}{*}{Mistral-7B} & H1 & Prior & 0.20 [$-$0.03, 0.43] & .07 \\
   & H2 & Closeness & $-$0.38 [$-$0.61, $-$0.15] & .21 \\
   & H3 & Hindsight & 0.39 [0.12, 0.66] & .16 \\
  \midrule
  \multirow{3}{*}{Qwen2.5-7B} & H1 & Prior & $-$0.00 [$-$0.31, 0.28] & .02 \\
   & H2 & Closeness & $-$0.26 [$-$0.55, 0.05] & .08 \\
   & H3 & Hindsight & $-$0.09 [$-$0.42, 0.24] & .03 \\
  \bottomrule
  \end{tabular}%
  }
  \end{table}

\subsection{Methods}
\subsubsection{Large Language Models}
We sampled responses from four LLMs: OpenAI's GPT-5.2 \citep{openai2025gpt52} (API identifier: \texttt{gpt-5.2}), accessed via the OpenAI API; Meta's Llama-3.1-70B-Instruct-Turbo \citep{grattafiori2024llama3herdmodels} (API identifier: \texttt{meta-llama/Meta-Llama-3.1-70B-Instruct-Turbo}), accessed via the Together AI API; Mistral AI's Mistral-7B \citep{jiang2023mistral} (identifier: \texttt{mistral:7b}); and Alibaba's Qwen2.5-7B \citep{qwen25technicalreport} (identifier: \texttt{qwen2.5:7b}), with the latter two run locally via Ollama.

\subsubsection{Procedure}
For each condition, we used the same 15 vignette-context trials and the same three response options as in Experiment~1. Each trial was presented as a prompt containing the vignette, the relevant question per condition, and three labeled options (A/B/C). LLMs were instructed to output only the chosen option letter. For each LLM $\times$ condition $\times$ vignette, we drew 50 samples using temperature $T=0.7$ and constrained generation to a single token so that the output was a single choice label. No conversational memory was used.

\begin{table}[b]
  \centering
  \caption{\textbf{Experiment 2 (LLMs)}. H4 combined linear model (foil $\sim$ hindsight $+$ prior $+$ closeness): $\beta$ [95\% CrI] per predictor, model $R^2$, and pairwise contrasts on $\Delta\beta$, by LLM.}
  \label{tab:exp2_h4}
  \resizebox{\columnwidth}{!}{%
  \small
  \begin{tabular}{@{}lcccc@{}}
  \toprule
   & GPT-5.2 & Llama-3.1-70B & Mistral-7B & Qwen2.5-7B \\
  \midrule
  Hindsight & 0.56 [0.31, 0.82] & 0.74 [0.39, 1.09] & 0.34 [0.08, 0.58] & 0.20 [$-$0.30, 0.70] \\
  Prior & $-$0.05 [$-$0.31, 0.22] & $-$0.04 [$-$0.50, 0.41] & 0.15 [$-$0.06, 0.34] & $-$0.06 [$-$0.40, 0.30] \\
  Closeness & $-$0.19 [$-$0.43, 0.07] & $-$0.27 [$-$0.68, 0.13] & $-$0.40 [$-$0.60, $-$0.19] & $-$0.37 [$-$0.78, 0.04] \\
  Model $R^2$ & .40 & .36 & .39 & .13\\
  \midrule
  \makecell[l]{$\Delta\beta$ \\ (hind $-$ prior)} & 0.61 [0.23, 1.00] & 0.78 [0.19, 1.38] & 0.19 [$-$0.11, 0.50] & 0.26 [$-$0.37, 0.88] \\
  \makecell[l]{$\Delta\beta$ \\ (hind $-$ close)} & 0.75 [0.45, 1.04] & 1.01 [0.53, 1.48] & 0.73 [0.46, 1.00] & 0.57 [$-$0.13, 1.27] \\
  \bottomrule
  \end{tabular}%
  }
\end{table}

\subsection{Results}
Data analyses for Experiment~2 mirror those of Experiment~1.

\subsubsection{Linear regression}
\begin{figure*}[t]
  \centering
  \includegraphics[width=\textwidth]{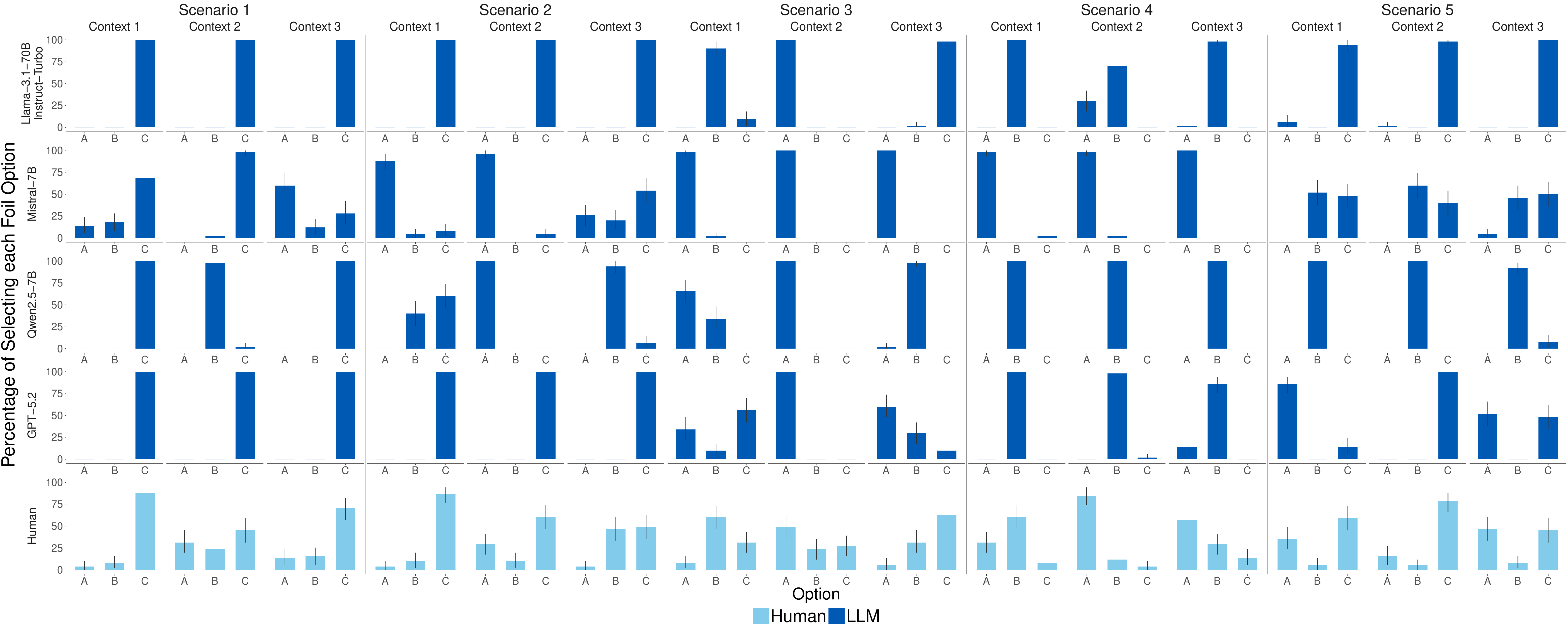}
  \caption{\textbf{Experiment 2 (LLMs)}. LLM and participants foil selections for all vignettes. Error bars are 95\% bootstrapped CIs.} 
  \label{fig:linear_regression_percentages_LLMs}
\end{figure*}

We performed linear regressions on vignette-level percentages (foil $\sim$ prior (H1); foil $\sim$ closeness (H2); foil $\sim$ hindsight (H3); combined model for H4) for each LLM to test if our hypotheses hold for LLMs as well. \autoref{tab:exp2_primary} summarizes $\beta$ (95\% CrI) and $R^2$ for H1--H3. The results for H4 for each of the models are reported in \autoref{tab:exp2_h4}. \autoref{fig:linear_regression_percentages_LLMs} shows the foil selection percentages for each LLM and for human participants.

\subsubsection{Multinomial prediction}
We again performed a Bayesian multinomial analysis on each LLM's foil choices (prior, closeness, hindsight models vs.\ uniform baseline).
\autoref{tab:exp2_llm_combined} reports $\Delta\text{ELPD}$ and SE by predictor and LLM. For all four LLMs, no predictor beat the uniform baseline (all $\Delta\text{ELPD}$ negative); within each LLM, hindsight had the least negative $\Delta\text{ELPD}$.

\subsubsection{Human--LLM similarity}
To compare each LLM's foil-selection distribution to that from the human participants, we computed cross-entropy $H(\text{human}, \text{LLM})$ per vignette: the human distribution was the empirical proportion of foil choices, and the LLM distribution was Dirichlet-smoothed ($\alpha = 0.5$) for numerical stability from the 50 samples per vignette. Lower mean cross-entropy indicates a more human-like foil distribution. We also report modal-match rates (out of 15 vignettes within a condition) for more intuitive interpretability.

\autoref{tab:exp2_llm_combined} reports mean cross-entropy and 95\% CIs per LLM. 
The modal-match analysis for foil selection resulted in Llama-3.1-70B 12/15 (80\%), GPT-5.2 10/15 (67\%), Mistral-7B 6/15 (40\%), and Qwen2.5-7B 5/15 (33\%); and, for hindsight, GPT-5.2 and Llama-3.1-70B 11/15 (73\%) each, Qwen2.5-7B 6/15 (40\%), and Mistral-7B 4/15 (27\%). Similarity metrics for all conditions can be found in the online materials.

\begin{table}[b]
  \centering
  \caption{\textbf{Experiment 2 (LLMs)}. Multinomial prediction performance on foil test set ($\Delta\text{ELPD}$ vs.\ uniform; SE in parentheses) and $H(\text{human}, \text{LLM})$ in nats with 95\% bootstrap CIs.}
  \label{tab:exp2_llm_combined}
  \resizebox{\columnwidth}{!}{%
  \begin{tabular}{@{}lcccc@{}}
  \toprule
   & Llama-70B & GPT-5.2 & Mistral-7B & Qwen-7B \\
  \midrule
  \multicolumn{5}{l}{\textit{Multinomial prediction: $\Delta\text{ELPD}$ (SE) vs.\ uniform}} \\[2pt]
  Prior      & $-$917.48 (73.66)  & $-$918.26 (72.40)   & $-$1910.37 (102.21) & $-$2084.32 (80.92) \\
  Closeness  & $-$959.43 (62.23)  & $-$1801.87 (66.14)  & $-$2778.66 (52.51)  & $-$2362.94 (63.72) \\
  Hindsight  & $-$281.47 (63.73)  & $-$442.82 (70.30)   & $-$327.08 (50.24)   & $-$1757.40 (75.93) \\
  \midrule
  \multicolumn{5}{l}{\textit{Human--LLM similarity: cross-entropy $H(\text{human}, \text{LLM})^*$}} \\[2pt]
  \makecell[l]{Mean CE\\ ([95\% CI]} & 1.65 [1.29, 2.00] & 1.75 [1.35, 2.22] & 2.10 [1.61, 2.62] & 2.63 [2.01, 3.22] \\
  \bottomrule
  \end{tabular}}
  \par\vspace{4pt}
  \begin{minipage}{\columnwidth}
  \scriptsize $^\ast$Human split-half baseline: 0.86 [0.79, 0.95].
  \end{minipage}
\end{table}

\subsection{Discussion}

Across LLMs, the linear regression results provided mixed support for the hypotheses: GPT-5.2 and Llama-3.1-70B supported H1 and H3 (but not H2), Mistral-7B supported H3 only, and Qwen2.5-7B supported none. In the combined models (H4), hindsight showed the clearest dominance for GPT-5.2 and Llama-3.1-70B, while evidence was weaker for Mistral-7B and absent for Qwen2.5-7B. However, the multinomial analysis did not find any predictor that outperformed a uniform baseline. Among the tested LLMs, Llama-3.1-70B and GPT-5.2 had the lowest mean cross-entropy to human foil selection and produced foil distributions most similar to humans, suggesting partial alignment with the pragmatic structure of foil inferences. Taken together, these results suggest that while some models produce foil choices that look consistent with hindsight in aggregate, their explicit expectation judgments are not reliably coupled.
The LLMs' choices follow high-probability token sequences that seem to overlap with human hindsight patterns, without being functionally grounded in them. The divergence between humans and LLMs, firstly, supports findings from Experiment~1 that foil selection is not just about choosing the most likely alternative. Secondly, this implies that LLMs struggle with inferring the implicit meaning of why-questions when the foil is unstated.

\section{General discussion}

How do we figure out what implicit contrast someone had in mind when they asked a why-question? The experiments reported here support a view of foil inference as a social inference problem: explainers infer which alternative contrast the questioner had in mind by reasoning about what becomes salient once the outcome is known. 

In future work we want to account explicitly for an explanation as a cooperative communicative act \citep{vanfraassen1988,HiltonErb1996MentalModels}. Current computational models formalize explanation choice as utility-driven communication under epistemic and social constraints \citep{chandra2024,goodman2016, KirfelIcardGerstenberg2022InferenceFromExplanation}. Thus, explainers may preferentially select foils for which they can provide a coherent explanation, biasing foil inference toward answerable contrasts. Therefore, explicitly manipulating what the questioner believes, and what the explainer believes -- both about what happened and about what the questioner believes -- will help us better understand the cognitive and social mechanisms underlying foil inference. 

\textcite{KominskyPhillips2019ImmoralProfessors} argue that explanatory reasoning privileges norm-conforming counterfactual alternatives over equally possible but norm-violating, causally remote, or downstream alternatives, such that explanations are structured around restoring what should have happened rather than exploring arbitrary ways the outcome could have differed. This might be a good starting point for further investigating people's foil inferences.

Our experiments provided participants with a discrete set of preselected foils to choose from. It would be interesting to conduct a similar experiment in a more naturalistic setting where participants freely generate foils. Work by \textcite{srinivasan2022shape} shows that in open-ended decision problems, people generate options via a structured search: early options are semantically clustered and high in value, while later options become increasingly dissimilar and lower in value. This suggests that option generation is guided by salience and learned normative structure rather than exhaustive exploration. 

Our experiments used a limited set of scenarios which cannot fully reflect the diversity of real-world situations. Moreover, in spoken conversation, the relevant contrasts are shaped by how the question is posed, including linguistic cues such as sentence stress, which can implicitly signal which part of the question should be contrasted \citep{Schaffer2005ContrastiveCausation}. At present, our design is text-based and does not capture the natural conversational context.

We focused here on why-questions about an agent's action in a particular situation. However, why-questions can be more general, and directed toward understanding how the world works (e.g., ``Why is the sun yellow?''; see \citealp{colombo2017}). More work is needed to better understand implicit contrasts in such cases.

The present work suggests that inferring the foil of a why-question is a form of social inference. People infer the intended contrast by reasoning about what the questioner likely expected before the outcome occurred, with hindsight expectation emerging as the strongest predictor of foil selection across scenarios and analyses. In contrast, large language models do not mirror human foil inference: different models show heterogeneous patterns, and their choices are not consistently aligned with their expectation judgments. Future work could use chain-of-thought elicitation to probe whether LLMs that select foils consistent with hindsight expectation genuinely reason about counterfactual alternatives and social inference. For now, this divergence suggests that in practical applications such as chatbot interactions, why-questions may require the intended foil to be stated explicitly.

\section{Acknowledgments}
We thank Lio Wong for helpful discussions. BB was supported by the Studienstiftung des deutschen Volkes (German Academic Scholarship Foundation) for her research stay at Stanford.
TG was supported by grants from the Stanford Institute for Human-Centered Artificial Intelligence (HAI) and from the Cooperative AI Foundation.
\printbibliography 

\end{document}